\documentclass[aps,prl,twocolumn]{revtex4}
\usepackage[T1]{fontenc}
\usepackage{amsmath}
\usepackage{amssymb}
\usepackage{graphicx}
\usepackage{setspace}
\usepackage{wasysym}
\usepackage{esint}
\usepackage{amsmath,bm}
\usepackage{color}
\usepackage{enumitem}

\begin{document}
\global\long\def\real#1{\mathbb{R}^{#1}}
\global\long\def\trace{\text{trace}}
\global\long\def\del{\nabla}
\global\long\def\cross{\times}
\global\long\def\diff#1#2{\frac{\partial#1}{\partial#2}}
\global\long\def\rot{\del\cross}
\global\long\def\div{\text{div}}

\title{{Freely flowing currents and electric field expulsion in viscous
electronics}}

\author{Michal Shavit, Andrey Shytov and Gregory Falkovich }
\affiliation{Weizmann Institute of Science, Rehovot 76100 Israel \\ University of Exeter, Stocker Road, Exeter EX4 4QL, UK}

\begin{abstract}
{Electronic fluids bring into hydrodynamics a new setting: equipotential flow sources embedded
inside the fluid. Here we show that nonlocal relation  between current and electric field due to
momentum-conserving inter-particle collisions leads to a total or partial field expulsion from such flows. That results in  freely flowing currents in the bulk and boundary jump in electric potential at current-injecting electrodes.   We derive the appropriate boundary
    conditions, analyze current distribution in free flows,
    discuss how the field expulsion  depends
    upon geometry of the electrode, and link the phenomenon to breakdown
    of conformal invariance.
  }
\vskip 0.2truecm
\end{abstract}

\maketitle


We experience now a rare moment of  {intense} 
interaction between fields of solids and fluids. This  is due to appearance
of new high-mobility materials where current carriers exchange momentum
faster than loose it to the lattice, so that their collective motion
is a viscous fluid flow
\cite{damle97,muller2009,sheehy2007,fritz2008,andreev2011,forcella2014,tomadin2014,narozhny2015,
  cortijo2015,lucas2016,bandurin2015, crossno2016,moll2016,Onset}.
Ideas from fluid mechanics can solve problems of nanoscale
  electronics:
in particular,   decrease resistance
below ballistic limit and make current flow against the electric
field \cite{bandurin2015,key-1,H2,KrishnaKumar2017}.
No less remarkable
is what electronics can do for fluid mechanics: 150 years after Stokes it
can  reveal new fundamental phenomena in laminar viscous flows, which is
the subject of this Letter. The reason is that   electronics brings
a new setting which had not been regularly considered in low-Reynolds
hydrodynamics --- equipotential (metallic) electrodes serving as flow sources embedded
inside the fluid.
We show below that the conditions on the electric
potential (pressure) imposed by sources  could be in conflict with those
of a viscous flow, which leads to anomalies at the boundaries and novel
flow properties, see Figure 1.

\begin{figure}
\begin{singlespace}
\includegraphics[scale=0.26]{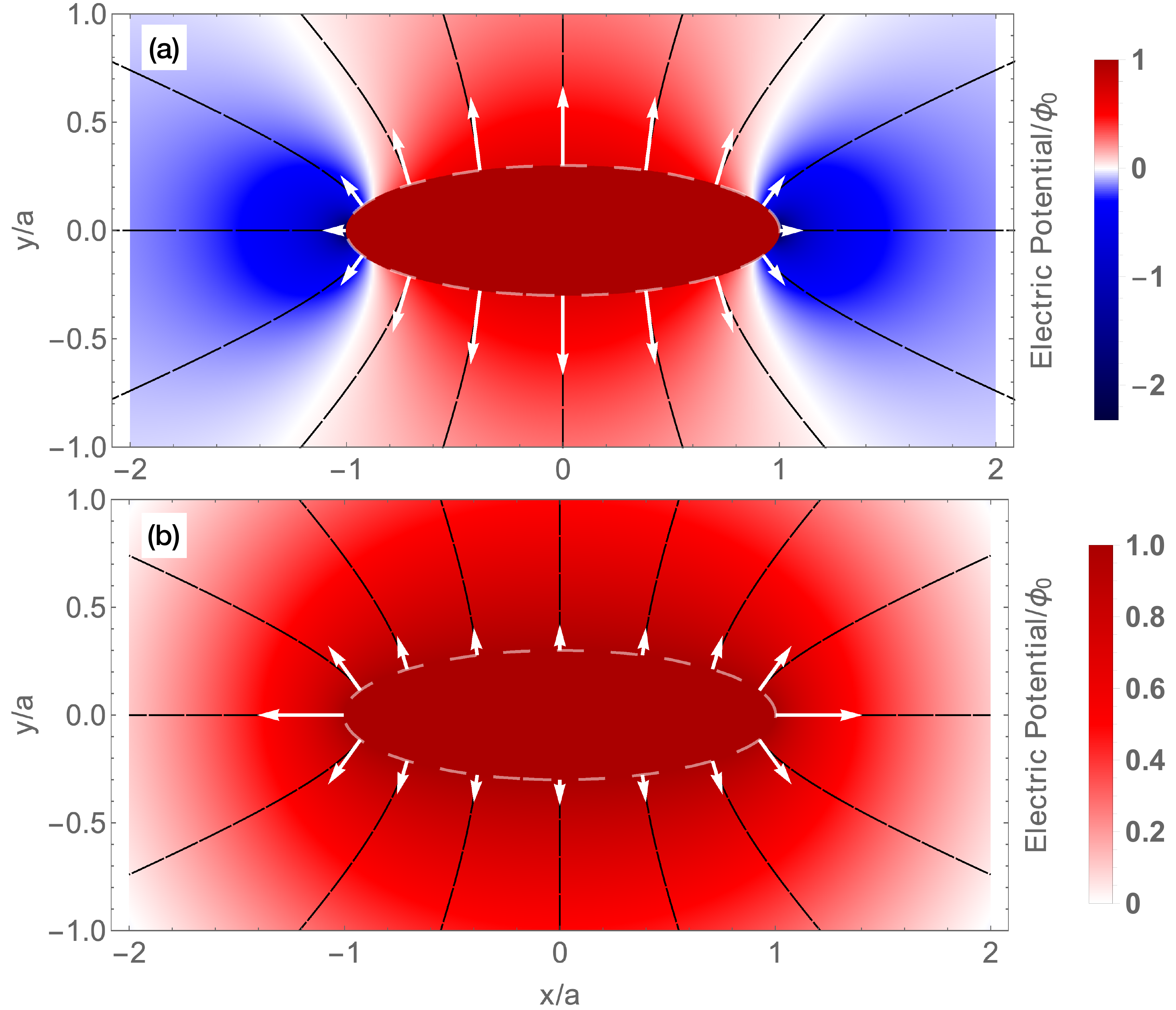}
\end{singlespace}
\caption{
  \label{fig:elliptic}
  Streamlines (black), velocity (white arrows) and potential color map
  for current from an elliptic source  {($b=0.3a$)}
  in  a) viscous and b) Ohmic regimes.
   {The current near the tips is enhanced for the Ohmic and
  suppressed for the viscous flow.}
   {Viscous drag near the tips results in a negative voltage whose magnitude exceeds the driving voltage .}
}
\end{figure}

Electronic fluids (e-fluids) are characterized by an unusual
response of electric current
to electric field~$\mathbf{E}$.
Instead of the usual Ohm's law, $ne \mathbf{v}  = \sigma \mathbf{E}$,
charge flows  at  the scales exceeding the electron-electron (e-e) mean free path $ l_{ee}$ are described by the combined Ohm-Stokes equation,
stating that the electric
field must now overcome both ohmic and viscous friction:
\begin{align}
  \left(\eta\Delta- n^2e^{2}/\sigma \right)\mathbf{v} & = - ne \mathbf{E}
  \label{OS}
\ .
\end{align}
Here $n,e,\mathbf{v}$ are the number carrier density, charge
and mean velocity, and $\sigma$  is the medium conductivity. The nonlocal first term in (\ref{OS}) represents momentum diffusion due to momentum-conserving
  scattering between charge carriers and is proportional to the viscosity~$\eta$.

Equation (\ref{OS}) presents an interesting puzzle.
  Consider an incompressible flow,
$\nabla\cdot \mathbf{v}=0$,
 and a potential electric field: $\mathbf{E} = - \del \phi$.
In that case, solving Eq.(\ref{OS}) appears deceptively simple:
 any solution with~$\eta = 0$ (purely Ohmic flow)
  also provides a solution for~$\eta \neq 0$,
since the viscous force vanishes identically:
$\Delta\mathbf{v} \propto \nabla\Delta\phi=0$.
This suggests a paradoxical  conclusion:
  the flow pattern~$\mathbf{v}(\mathbf{r})$ is not affected by viscous
  friction, and the electric potential driving the flow vanishes
  in the purely viscous limit~$\sigma \to \infty$.

Vanishing of the viscous force, however counterintuitive,
can be verified explicitly in the simplest case of a
spherical metallic electrode ejecting radial current into an infinite medium.
The velocity field 
is readily obtained from current conservation:
\begin{align}
  ne\mathbf{v}\left(\mathbf{r}\right) & =I \mathbf{e}_r/\Omega_d r^{d-1}\ .
                                        \label{eq:Circ flow}
\end{align}
Here $ {r}$ is the distance to the source center, $I$ is the total current,
$\Omega_d$
is the area of a unit sphere
in $d$ dimensions.
Since \textbf{$\mathbf{v}$}
is a gradient of a harmonic function, the viscous force $\Delta\mathbf{v}$
vanishes everywhere
(this remains true if one adds uniform circulation, for instance, $v_\theta\propto 1/r$ in 2d). The  electric field outside
is  indeed independent of viscosity: $\del\phi=I\mathbf{e}_r/\Omega_d r^{d-1}\sigma$.
We therefore conclude that electric field  is not required to drive
a purely viscous flow within the system bulk. Yet the paradox remains:
 the viscous stress tensor,
$\sigma_{ij}=\eta\left(\partial_{j}v_{i}+\partial_{i}v_{j}\right) $, is nonzero as well as
 the energy dissipation rate due to viscous friction,
 \begin{equation}
   P=  {\frac{1}{2\eta}}
   \int\sum_{i,j}\left(\sigma_{ij}\right)^{2}dV \,.\label{P}\end{equation}
In other words, even though the  {net} viscous
force (divergence of the stress tensor) acting on any fluid element
is zero, there are nonzero forces acting on the opposite sides of
the element and deforming it, which must lead to dissipation.
The energy loss~$P$ in the bulk must be compensated
by the  {work $\phi_0 I$ performed by the current} source.
 This requires a finite electrode potential relative to sink at infinity:
$\phi_0 =P/I= 2\eta\left(d-1\right)I/\left(ne\right)^{2}\Omega_{d}a^{d}$ ($a$ is the electrode radius).
The contribution of the infinite viscous medium to the total resistance
of the system is thus determined by the electrode size:
$R =\phi_0/I=2\eta\left(d-1\right)/\left(ne\right)^{2}\Omega_{d}a^{d}$.
 {To reconcile finite~$\phi_0$
  with vanishing~$\phi({\bf r})$ in the bulk,
  we conclude that the potential distribution must exhibit
a sharp viscosity-dependent drop}
across  {a thin} Knudsen layer of  width $\sim l_{ee}$,
 where the Stokes equation is not applicable. Note that the jump of the potential (pressure) at the boundary gives the momentum flux
$(ne) \phi_0$, which is exactly equal to the normal component of the viscous stress tensor,
$\sigma_{nn} = - 2 \eta \partial_r v_r$. In other words, the potential discontinuity  {can be paraphrased as} the continuity of the normal flux
of normal momentum. Indeed, viscous stresses are absent inside the electrode
but present outside, so  that the potential jump  compensates the jump
in the stress.
That potential jump is similar to the  Kapitza temperature jump upon
heat transfer through a solid-liquid interface~\cite{ref:Kapitza-jump}.

The electric field is thus
expelled from the bulk and is concentrated in the boundary layer
in a viscous flow.
Potential jump is proportional to the viscosity, that is to the mean free path. That means that the electric field inside the ballistic layer is independent of the mean free path.  When one goes deep into the hydrodynamic regime (say, by increasing temperature in graphene), the mean free path shrinks but the electric field stays finite.

Stokes encountered similar phenomenon of bulk dissipation equal to the surface work in his analysis
  of the decay of water waves: the flow in the bulk is
  potential, while the viscous forces
  only perform work on the surface \cite{key-9}, see also \cite{key-3}.

Expulsion of the field from the bulk and its concentration at microscopic scales 
  can be verified by analyzing kinetics of
  momentum-conserving e-e collisions \cite{key-5,key-4} for a point-like
  electrode, $a\ll l_{ee}$. Neglecting Ohmic momentum losses,
  such approach yields a potential that  in the ballistic domain $r \ll l_{ee}$ decays slowly:
 $\phi({\mathbf r}) \propto 1/r$ in~$d=2$. The potential falls rapidly at large distances,
  $\phi(\mathbf{r}) \propto \exp(-r/l_{ee})$.
In other words, a point source produces a radial flow having a constant
potential at $r\gg l_{ee}$.
 By superposition this is also true for an arbitrary combination of point sources and sinks. However,
field expulsion is only approximate for finite-size sources and sinks
because they impose boundary conditions.

How is the above picture of field expulsion modified for an electrode of an
arbitrary shape? A purely Ohmic flow in two dimensions
can be found via conformal mapping
that deforms one electrode into another and also transforms
stream lines and potential contours.
Naive reasoning  {outlined above}
suggests that the transformation of the potential might also be possible when viscosity is present due to conformal invariance of the Laplacian.
Below we demonstrate that this conformal  {equivalence}
does not hold for viscous flows, since the field distribution
depends nontrivially upon the shape of the electrode;
in particular, for non-symmetric electrodes the flow, in general, is not potential and electric field partially penetrates the fluid.

\begin{figure}[h]
\begin{raggedright}
\includegraphics[scale=0.245]{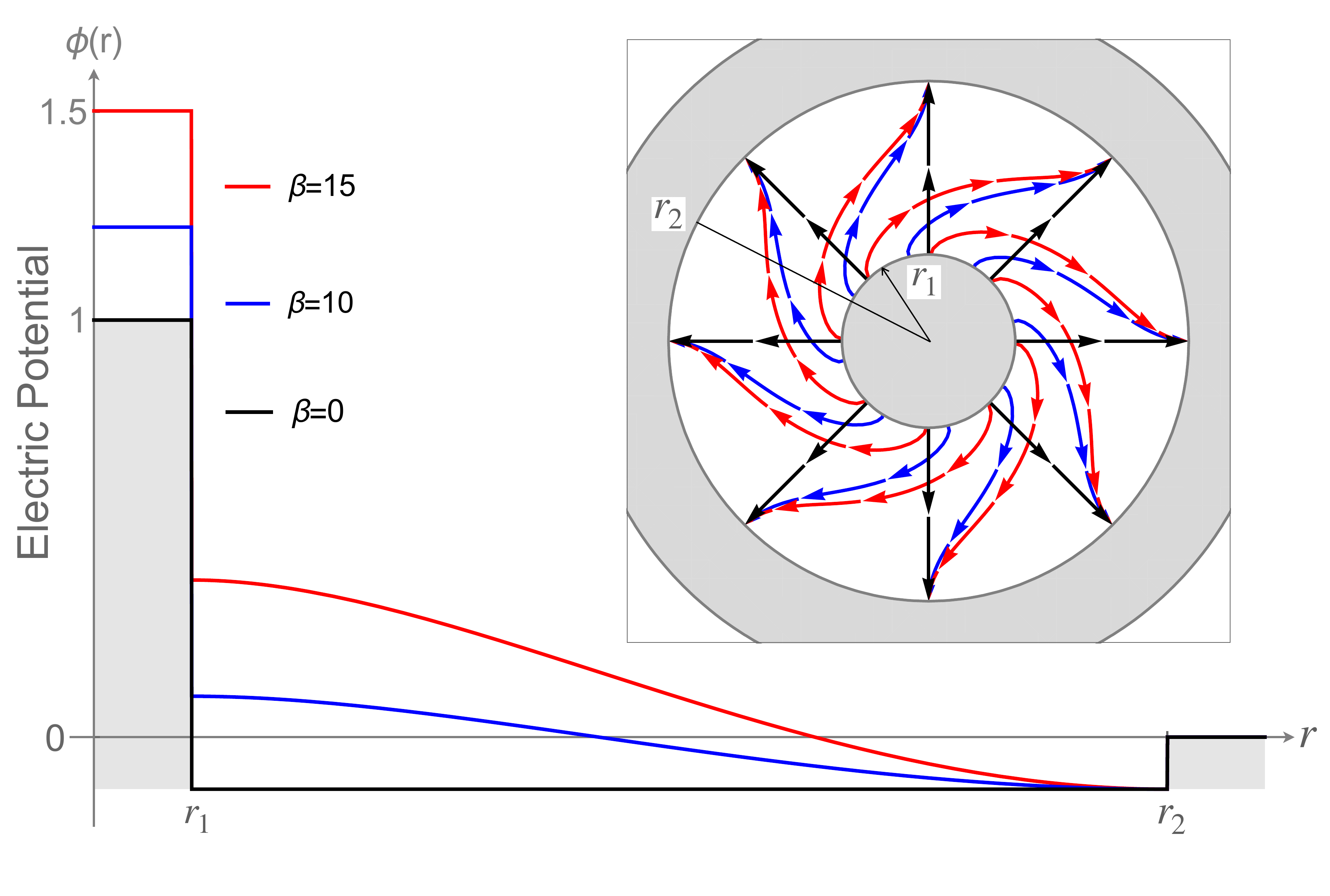}
\par\end{raggedright}
\caption{\label{fig:B circs}Potential (central plot, normalized to source
potential value) 
and streamlines (right box, normalized to the radial velocity on the source) for current flowing between two concentric
circular electrodes,  {$r_2 = 3r_1$}.
Different colors indicate different values of
$\beta$. The black line shows the potential jumps for  $\beta= 0$, when the electric field is concentrated on the electrodes and is zero in the bulk.
}
\end{figure}

Indeed, to determine the flow, one needs to
solve Eq.~(\ref{OS}) supplemented with boundary
conditions, which must follow from the same variational principle that gives  Eq.~(\ref{OS}),  i.e. minimization of the dissipated energy (\ref{P}).
Let us consider the dissipation rate~$P[\mathbf{v}]$ as a functional
of the velocity field~$\mathbf{v}(\mathbf{r})$, and minimize
it for a given total current emitted by the electrode. Variation with respect to the bulk velocity gives (\ref{OS}), while variation with respect to the normal velocity on the source
confirms the boundary condition (b.c.) in the form of the normal flux continuity (see Supplement):
\begin{align}
  ne\left[\phi_0 - \phi(\mathbf{r})\right]
  =- \sigma_{ij} n_i n_j \equiv \sigma_{nn} \ .\label{bc0}
\end{align}
where~$\mathbf{n}$ is the unit vector normal to the boundary and $\phi(\mathbf{r})$ is the boundary value of the potential satisfying Eq.(\ref{OS}).
The second boundary condition depends on the nature
of the interface between the source and the fluid.
In particular, one can consider  either a
 momentum-relaxing no-slip interface
with $\mathbf{v}_t =  \mathbf{v} \times \hat{\mathbf{n}}= 0$,
or  a smooth no-stress interface with $\sigma_{nt} = 0$.
In what follows, we restrict ourselves to the no-slip case for simplicity. In this case,
using incompressibility~$\nabla \cdot \mathbf{v} = 0$, one can rewrite (\ref{bc0}) via
the signed extrinsic curvature~$K$ of the boundary:
\begin{align}
  \label{eq:Pot drop curv}
  ne\left[\phi_0 - \phi(\mathbf{r})\right]
  =  2 \eta K (\mathbf{v} \cdot \hat{\mathbf{n}}) \ .
\end{align}
This can be interpreted as a universal, viscosity-dependent contribution
to contact resistance which cannot be ignored for a non-flat electrode.
Let us stress the remarkable fact that this contribution can be both positive and negative depending on the curvature sign. Indeed, to balance the viscous stress, electric field in the ballistic layer is directed along/against the current respectively for positive/negative curvature. For example, consider the viscous flow in an annulus between
two concentric circular electrodes
of radii  {$r_1$} and   {$r_2> r_1$},
known as the Corbino disc geometry (see Fig.~\ref{fig:B circs}).
The resistance of such system is determined by the
two potential jumps each given by Eq.~(\ref{eq:Pot drop curv}):
  {
  \begin{equation}
    R_0 = \frac{\eta}{\pi (ne)^2} \left(\frac{1}{r_1^2} - \frac{1}{r_2^2}\right)
    \ .\label{Corb1}
  \end{equation}
 }
Since the Stokes equation is symmetric  {under}
$\mathbf{v}\rightarrow-\mathbf{v}$, $\phi\rightarrow-\phi$,
 {reversal} of the current reverses the jumps.

If the curvature~$K$ varies along the interface,
the potential~$\phi(\mathbf{r})$, in general, is not a constant, which
results in a nonvanishing field in the bulk.
Therefore, Eq.(\ref{bc0}) provides an example of conformal anomaly
in classical physics: phenomena at microscopic length scale~$l_{ee}$
affect the flow at however larger distances in a universal way,
invalidating conformal invariant solutions to Eq.(\ref{OS}).

To see the breakdown of conformal invariance and partial penetration of the field into the flow, we consider an ellipse as the simplest non-trivial example of a source with a variable curvature.
 We introduce
elliptic coordinates~$\rho, \theta$:
\begin{align*}
  &x =\sqrt{a^{2}-b^{2}} \cosh\rho\cos\theta,\  y =\sqrt{a^{2}-b^{2}} \sinh\rho\sin\theta,
\end{align*}
where~$a > b$ are the semiaxes, $0 \leq \theta \leq 2\pi$ is the polar
angle, and~$\rho \geq \rho_0 = \tanh^{-1}(b/a)$ is the radial variable,
$\rho = \rho_0$ at the electrode.
This yields orthogonal coordinates in which
the scaling factors~$h_i = |\partial \mathbf{x} / \partial i|$
are equal:
\begin{align*}
  h_{\rho} = h_{\theta} =
  \sqrt{\left(a^{2}-b^{2}\right)\left(\sinh^{2}\rho+\sin^{2}\theta\right)},
\end{align*}
so that the variables~$(\rho, \theta)$ are related to~$(x, y)$ by
a conformal map, which facilitates the calculation of Laplacians.
We seek
solution of the Stokes equation, that is  (\ref{OS}) with $\sigma=\infty$,
supplied with the b.c.~(\ref{eq:Pot drop curv}) with the
curvature
$K = (h_\rho h_\theta)^{-1} \partial_\rho h_\theta$.
We also assume zero tangential
velocity~$v_\theta = 0$ at~$\rho = \rho_0$,
electrode   potential~$\phi_0$
and zero potential at infinity. Interestingly, the  velocity field  is  radial
everywhere
(the details are in the Supplement):
\begin{align}
\mathbf{v}\left(\rho,\theta\right) &
       = \frac{ne\phi_{0} (a^2 - b^2)}{\eta h_\theta(\rho, \theta)}
           (\sin^{2}\theta + \sinh^{2}\rho_{0})
          \, \mathbf{e}_\rho
\ .
\label{eq:elliptic}
\end{align}
The potential outside the source is non-uniform,
\begin{align}
  \frac{\phi\left(\rho,\theta\right)}{\phi_{0}}= 1-
     \frac{\sinh2\rho}{2\left(\sin^{2}\theta+\sinh^{2}\rho\right)} \,,
\label{pot:elliptic}
\end{align}
which gives non-vanishing electrical field inside the viscous domain.
At  large distances, the potential
is  a quadrupole proportional to the eccentricity of the
ellipse: $\phi(x, y) \propto (a^2 - b^2)  {(y^2 - x^2)}/r^4$.
It changes sign on the lines $x=y$, as for a point source in
a half-plane \cite{key-2}.

It is instructive to compare the viscous flow (\ref{eq:elliptic}) to the
purely ohmic flow,
${\bf v}_\Omega(\rho, \theta) \propto h_\theta^{-1} \hat{\mathbf{e}}_\rho$,
$\phi_\Omega (\rho, \theta) \propto \rho$, which can be obtained
by conformal deformation of the flow emitted by a circular electrode. The comparison can be seen in Fig.~\ref{fig:elliptic}.
The ohmic flow is also radial, with current concentrated near the tips
$\theta = 0, \pi$, where the curvature is maximal.
In a sharp  contrast, viscous current
mainly flows from flatter parts, while  at the tips it
is  suppressed by the~$\sin^2\theta$
factor in (\ref{eq:elliptic}).
The slow viscous current along the directions around the minimum is dragged
by the viscous force from adjacent faster currents. That viscous force is balanced by the electric
field  directed against the current \cite{key-1}. The potential jump at the electrode tip, according to (\ref{pot:elliptic}), is $-\phi_0a/b$, which for sufficiently eccentric ellipse can {significantly}
exceed, by absolute magnitude, the driving voltage {$\phi_0$}.

The conformal invariance is  indeed broken:
the solution (\ref{eq:elliptic}-\ref{pot:elliptic}) cannot be obtained
from the flow outside a circular electrode by a conformal mapping.
Since (\ref{eq:elliptic}) satisfies both the
no-slip and no-stress boundary conditions on the source,
our conclusions here are quite general.

The  resistance of the medium  depends on the source shape and size.
In the elliptic case, one finds from Eq.(\ref{eq:elliptic}):
\begin{align}
  R=\frac{\phi_{0}}{I}
  & =\frac{2\eta}{\pi\left(ne\right)^{2}\left(a^{2}+b^{2}\right)}\,.
\end{align}
The limit $b\rightarrow a$ reproduces the  resistance
for a circular source $R = \eta /\pi\left(nea\right)^{2}$. Comparing that with the kinetic regime, where $\phi\propto 1/r$,  we see that the voltage and resistance grow with decreasing
the electrode size as $1/a^{2}$ for $a\gg l_{ee}$ and as $1/a$
for $a\ll l_{ee}$, according to the general relation between viscous
and ballistic regimes \cite{key-1,key-2}. The resistance of a
  flat electrode, $b \ll a$, is determined
  by its width~$a$: $R = 2\eta / \pi (nea)^2$.

Stokes equation together with the charge continuity relates~$\mathbf{E}$
to vorticity~$\omega \equiv \nabla\cross\mathbf{v}$ of the flow:
\mbox{$ne\mathbf{E}=\eta\nabla\cross\omega$}.
Hence penetration of the electric field into the fluid
makes the flow non-potential.
One can quantify the degree of field
expulsion by the dimensionless parameter
\begin{align}
  \xi \equiv  {1 -}  {\eta\int\omega^{2}dV}/P\ ,
  \qquad 0 \leq \xi \leq 1 ,\label{xi}
\end{align}
 {so that work~$\xi e\phi_0$ is performed per each particle
crossing the Knudsen layer.}
For an elliptic source,
$\xi$ takes a particularly transparent form:
 {$\xi  = {2ab}/\left({a^{2}+b^{2}}\right)$}.
In particular, $\xi = 1$ for a circular source, $b=a$,
 when the field is fully expelled from the fluid,
 and $\xi = 0$  for a flat source,
 $b\ll a$, when the potential jump vanishes according
 to (\ref{eq:Pot drop curv}).

So far, we considered  vorticity  generated by a non-uniform
current through the electrode boundary with a non-uniform curvature. Vorticity can be also generated
by non-potential forces, such as Lorentz or Coriolis force.
Applying magnetic field~$B$ results in $B$-dependent velocity and
induces electric field in the bulk already in the circularly symmetric geometry. Consider the above Corbino geometry with two concentric electrodes.
Adding the magnetic field, we account only for the Lorentz force and disregard Hall viscosity
assuming large enough scales \cite{Hall1,Hall2}.
 The Lorentz force gives rise to an angular velocity and vorticity
and generates the electric field in the bulk. The force acting
on the angular current affects the potential drop and the resistance:
 \begin{align} R(B) = R_0\,+\,
       \frac{B^2 r_2^2}{16\pi\eta}
          \left(1 - \frac{4 \gamma ^{2} \ln^{2}\!\gamma  }{
                         \left(\gamma ^{2}-1\right)^{2}}
          \right)
\label{eq: R(B)}
\end{align}
where $R_0$ is given by (\ref{Corb1}),
 {and~$\gamma \equiv r_2 / r_1$ is the
aspect ratio}. The second term in the rhs of (\ref{eq: R(B)})
is the magnetoresistance, which is positive. The magnetoresistance
 {quickly}
grows with $r_2$ and is inversely proportional to $\eta\propto  l_{ee}$,
so that it  may easily exceed the boundary contribution   as the system
goes deeper into the fluidity-dominated regime.
 Redistribution of the field between boundary and bulk
is characterized by a dimensionless
number~$\beta \equiv ne B {r_2^2} / \eta$, which
is the ratio of Lorentz and  viscous forces at the
  outer rim and determines the number of
  turns the flow makes between the source and the sink.
Fig. \ref{fig:B circs}  illustrates the dependence
of the field inside the bulk for a varying $\beta$ and its expulsion
at $\beta=0$.

Note briefly that nonzero Ohmic resistivity always dominates at sufficiently large distances for zero magnetic field; to see expulsion  we need~$l_{ee} \ll a \ll l_*$,
where~$l_* = \sqrt{\sigma \eta} / ne$
is the ohmic-to-viscous crossover scale \cite{key-2}.
For the Ohmic-Stokes flow between the concentric electrodes, extra resistance  $(2\pi \sigma)^{-1} \log(r_2/r_1)$ is added to (\ref{Corb1}).
The viscous
  term saturates at~$r_2 \to \infty$, while  Ohmic  resistance
 {slowly} grows with  {increasing $r_2$} and dominates when
 {$r_1^2  \ln(r_2/r_1) \gg {(en)^{2}/\eta\sigma} $}.

While both Ohmic and viscous flows are inherently dissipative, there
is a dramatic 
difference in the spatial
distribution of the work done to compensate this dissipation. In Ohmic
flows the momentum and the energy losses are locally compensated by
an electric field proportional to the current at every point. On the
contrary, momentum is diffusively conserved by viscous flows while
the energy is lost everywhere there is a velocity gradient. As we have
shown here, the electrical work compensating the viscous energy loss can be
partially or even fully done on the flow boundaries.
Although we used the electronic terminology, all the statements are valid for a general incompressible
viscous fluid via the replacement
$ne \phi\rightarrow p$, with $p$ being the pressure.

Distributions of potential and current described here
could be probed e.g. in graphene via nanoscale imaging
methods,  {such as scanning-gate microscopy}~\cite{ref:Ilani-imaging, ref:Ensslin-SGM}. Instead of
embedding isolated contacts, it may be more practical
to use needle-shaped electrodes protruding into e-fluid from its edge.
Most of our findings remain valid for such a setup: due to~(\ref{bc0}), the
current is redistributed similarly near the tip, and viscous drag
results in 
a strong negative potential.  {Note} that the flow could be affected by a non-universal contact resistance~$R_c$.  {Still},
the viscous flow remains qualitatively distinct from an ohmic flow even when $R_c$ exceeds the resistance of the medium~$R$ ( {see the Supplement for an example}). Indeed, large~$R_c$ affects current uniformly along the contact, so that ``windshear'' near the tip may still induce negative
potential. Finally, enhancement of viscous effects at the onset of fluidity
observed recently~\cite{Onset} suggests that the negative potential
near the tip may be maximal in this crossover regime, i.e.
when~$l_{ee} \sim b$. Detailed analysis of this regime is beyond the
scope of this article.

To conclude, we demonstrated how electric field or pressure gradient can be partially or even completely expelled from viscous flows, so that some or all work compensating viscous dissipation is done on the source. In the electronic setting, when the electrodes are equipotential, the field and the work are concentrated in the ballistic layer on the surface. In hydrodynamic setting (say for a vertical tube injecting fluid between horizontal plates) the pressure gradient and the work are distributed inside the source.
We have shown that whether the viscous flow is totally or partially force-free depends on the geometry, particularly on the curvature of the source.
\begin{acknowledgments}We are grateful to Leonid Levitov for numerous helpful
discussions; his input was indispensable for this work. GF thanks Howard Stone for a helpful advice. The work was supported by the grants from
the  Minerva Foundation with funding from the Federal German Ministry for Education and Research, the Israeli Science Foundation,  the Center for Scientific Excellence and
the Russian Science Foundation (Project No. 14-22-00259).
\end{acknowledgments}


\begin{thebibliography}{1}
\bibitem{damle97}
Damle, K. \& Sachdev,  S.
Nonzero-temperature transport near quantum critical points.
{\it Phys. Rev. B} 56: 8714-8733 (1997).

\bibitem{muller2009}
M\"uller, M., Schmalian,  J. \& Fritz, L.
Graphene: A Nearly Perfect Fluid.
{\it Phys. Rev. Lett.} 103:025301 (2009).


\bibitem{andreev2011}
Andreev, A. V., Kivelson, S. A. \& Spivak, B.
Hydrodynamic Description of Transport in Strongly Correlated Electron Systems.
{\it Phys. Rev. Lett.} 106:256804 (2011).

\bibitem{forcella2014}
Forcella, D., Zaanen,   J., Valentinis, D. \& van der Marel, D.
Electromagnetic properties of viscous charged fluids.
{\it Phys. Rev. B} 90:035143 (2014).

\bibitem{tomadin2014}
Tomadin, A., Vignale,  G. \& Polini, M.
Corbino Disk Viscometer for 2D Quantum Electron Liquids.
{\it Phys. Rev. Lett.} 113:235901 (2014).

\bibitem{sheehy2007}
Sheehy, D. E. \& Schmalian, J.
Quantum Critical Scaling in Graphene.
{\it Phys. Rev. Lett.}~99:226803 (2007).

\bibitem{fritz2008}
Fritz, L., Schmalian,  J., M\"uller,  M. \& Sachdev, S.
Quantum critical transport in clean graphene.
{\it Phys. Rev. B} 78:085416 (2008) .

\bibitem{narozhny2015}
Narozhny, B. N., Gornyi, I. V., Titov,  M., Sch\"utt, M. \& Mirlin, A. D.
Hydrodynamics in graphene: Linear-response transport.
{\it Phys. Rev. B} 91:035414 (2015).


\bibitem{cortijo2015}
Cortijo, A., Ferreir\'os,  Y., Landsteiner, K. \& Vozmediano, M. A. H.
Hall viscosity from elastic gauge fields in Dirac crystals.
{\it Phys. Rev. Lett.} 115:177202 (2015).

\bibitem{bandurin2015}
Bandurin, D. A. {\it et al.} 
Negative local resistance caused by viscous electron backflow in graphene.
{\it Science} 351:1055-1058 (2016).


\bibitem{crossno2016}
Crossno, J. {\it el.}
Observation of the Dirac fluid and the breakdown of the Wiedemann-Franz law in graphene.
{\it Science} 351:1058-1061 (2016).

\bibitem{moll2016}
Moll, P. J. W., Kushwaha,  P., Nandi,  N., Schmidt,  B. \& Mackenzie, A. P.
Evidence for hydrodynamic electron flow in ${\rm PdCoO_2}$.
{\it Science} 351:1061-1064 (2016).

\bibitem{Onset} D. Bandurin {\it et al}, Fluidity onset in graphene, Nature Communications {\bf 9}, 4533 (2018)

\bibitem{lucas2016}
Lucas, A., Crossno,  J., Fong,  K. C., Kim,  P. \& Sachdev, S.
Transport in inhomogeneous quantum critical fluids and in the Dirac fluid in graphene
{\it Phys. Rev. B} 93:075426 (2016).







\bibitem{H2}
Guo, H., Ilseven,  E.,  Falkovich,  G. \& Levitov, L.
Higher-than-ballistic conduction of viscous electron flows.
{\it Proc. Natl. Ac. Sci.} 114:3068-3073 (2017).

\bibitem{KrishnaKumar2017}
Krishna Kumar, R. {\it et al.} 
Superballistic flow of viscous electron fluid through graphene constrictions.
{\it Nature Phys.} 13: 1182  (2017) .

\bibitem{key-1} Levitov, L., and Falkovich, G.,  \textquotedbl{}Electron
viscosity, current vortices and negative nonlocal resistance in graphene.\textquotedbl{}
Nature Physics 12.7 (2016): 672.

\bibitem{ref:Kapitza-jump}
P.L.Kapitza, J.Phys (USSR) {\bf 4}, 181 (1941)

\bibitem{key-9}Stokes, G.G. (1845) On the Theories of the Internal
Friction of Fluids in Motion and of the Equilibrium and Motion of
Elastic Solids. Trans. Cambridge Philos. Soc., 8, 287-319.

\bibitem{key-3} Falkovich, G., Fluid Mechanics (Cambridge Univ. Press
2018)

\bibitem{key-5} Guo, H.,  Ilseven, E.,  Falkovich, G.,
Levitov, L.S.,  Higher-than-ballistic conduction of viscous electron
flows. PNAS 114, 3068-3073 (2017)








\bibitem{key-4} Shytov, A., Kong, J.F., Falkovich, G.,
Levitov, L.,  Electron Collisions and Negative Nonlocal Response
of Ballistic Electrons. arXiv:1806.09538

\bibitem{key-2} Falkovich, G., \& Levitov, L. (2017). Linking spatial
distributions of potential and current in viscous electronics. Physical
review letters, 119(6), 066601. ISO 690






\bibitem{Hall1}T Scaffidi, N Nandi, B  Schmidt, A P  Mackenzie, and J  E Moore, Hydrodynamic Electron Flow and Hall Viscosity Phys. Rev. Lett. 118, 226601 (2017)

\bibitem{Hall2}
A. I. Berdyugin et al, 
arXiv:1806.01606

\bibitem{ref:Ensslin-SGM}
  Braem, B.A., Pellegrino, F.M.D., Principi, A., R\"o\"osli, M.,
  Hennel, S., Koski, J.V., Berl, M., Dietsche, W., Wegscheider, W.,
  Polini, M., Ihn, T., and Ensslin, K.,
  Scanning Gate Microscopy in a Viscous Electron Fluid,
  arXiv.org:1807.03177 (2018)

\bibitem{ref:Ilani-imaging}
  Ella, L., Rozen, A., Birkbeck, J., Ben-Shalom, M.,
  Perello, D., Zultak, J., Taniguchi, T., Watanabe, K.,
  Geim, A.K., Ilani, S., Sulpizio, J.A.
  arxiv.org:1810.10744 (2018)
\end{thebibliography}
\end{document}